# Elastic potentials as yield surfaces for homogeneous materials

Jorge Castro
University of Cantabria
Avda. de Los Castros, s/n
39005 Santander, Spain
e-mail: castrogj@unican.es

**Abstract**

This paper proposes that elastic potentials, which may be rigorously formulated using the negative Gibbs free energy or the complementary strain energy density, should be used as the basis for the plastic part of elasto-plastic constitutive models. Thus, the yield surface may be assumed as an elastic potential surface for a specific level of critical complementary strain energy density. Here, rate-independent homogenous continuous materials under isothermal conditions are considered. Visualization of elastic potentials using principal stresses is presented.

The proposed approach improves the total strain energy criterion because: (1) the elastic potential does not have to be centred at the current stress state and, consequently, is able to reproduce a tension-compression asymmetry; (2) the corresponding correlation between the Poisson's ratio and the shape of the yield surface is found for soils and metallic glasses; (3) non-linear elasticity is considered, which notably increases the flexibility and capabilities of the proposed approach.

Ultimately, and similarly to hyperelasticity, the proposed framework for deriving (associated) yield surfaces may be considered just as a classifying criterion and a possible approach to formulate yield surfaces. Finally, if an associated flow rule is also assumed, the elastic potential, yield and plastic potential surfaces coincide.

# 1. Introduction

Although the concepts of work and energy are essential in continuum solid mechanics, they are not so commonly or easily integrated in yield criteria. Notable attempts, such as the total strain energy criterion (Beltrami 1885; Haigh 1920), are not currently used. In the related field of fracture mechanics, energy is generally accepted as a criterion for crack initiation; in fact, its origin is due to Griffith (1921), who originally applied the first law of thermodynamics to solve the failure problem of a cracked glass and proposed a critical energy criterion. Nowadays, one of the successful methods for fracture assessment is the strain energy density method (e.g., Lazzarin and Zambardi 2001).

At the same time, energy concepts are helpful in providing additional techniques to solve elasticity problems (e.g., Sadd 2014). Also, a restrictive form of elasticity that is usually called hyperelasticity (e.g., Fung 1965) requires the existence of strain energy potential functions (Figure 1):

$$\sigma_{ij} = \frac{\partial U_0}{\partial \varepsilon_{ij}} \text{ and } \varepsilon_{ij} = \frac{\partial U_{c0}}{\partial \sigma_{ij}} \tag{1}$$

Cauchy stresses and small strains are considered in this paper.

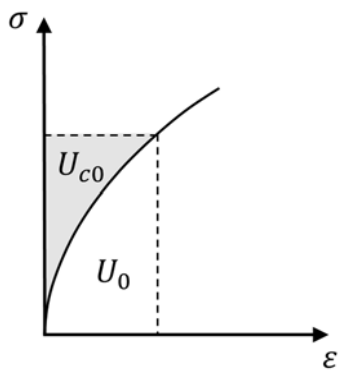

Figure 1. Strain energy for uniaxial stress.

Any kind of elasticity automatically satisfies the second law of thermodynamics because the stress-strain behaviour is reversible, but hyperelastic formulations also satisfy the first law of thermodynamics automatically. For isothermal conditions, the strain energy per unit volume, $U_0$, also called strain energy density, and the complementary strain energy per unit volume, $U_{c0}$, are equivalent to the Helmoltz free energy and Gibbs free energy with negative sign, respectively (e.g., Houlsby and Puzrin 2006).

Motivation for the present work emerges from the fact that the elastic limit (yield surface) should be related to the elastic behaviour and, for example, it seems logical to assume that the stiffness of a material increases with the mean pressure, as its yielding surface does, or vice versa.

This paper proposes that elastic potentials, which may be rigorously formulated using the negative Gibbs free energy or the complementary strain energy density, should be used as the basis for the plastic part of elasto-plastic constitutive models. Thus, the yield surface may be assumed as an elastic potential surface for a specific level of critical complementary strain energy density. Here, rate-independent homogenous continuous materials under isothermal conditions are considered. Section 2 presents the case of linear isotropic materials, both incompressible and compressible materials, where elastic potentials lead to von Mises and elliptical yield surfaces, respectively. Section 3 further examines non-linear materials, which provide distorted elliptical yield surfaces and, for the case of an incompressible material, could lead to Tresca criterion. Section 4 briefly introduces linear anisotropic elasticity, which leads to a rotation of the elastic potential.

Despite the intended generality, the sign convention and most of the examples reflect the author's bias for geomaterials. This paper does not intend to present "universal" yield criteria; rather, a theoretical framework within yield criteria may be formulated, and its potential capabilities, such as relating elastic and yielding parameters.

## 2. Linear isotropic elasticity

### 2.1 Elastic potential

Linear elasticity may be easily formulated within the hyperelastic framework (e.g., Sadd 2014); it is enough to assume that the elastic potential is a quadratic form:

$$U_{c0} = U_0 = a\sigma_i^2 + b\sigma_i\sigma_j \qquad i,j,k = 1,2,3 \tag{2}$$

Here, contracted notation is used for the sake of brevity and elastic potentials are presented in terms of unordered principal stresses, $\sigma_i$, for the sake of visualization in the principal stress space. As the material is isotropic, the behaviour for each principal direction should be identical. It is quickly demonstrated that:

$$\frac{\partial^2 U_{c0}}{\partial \sigma_i^2} = 2a = \frac{1}{E} \text{ and } \frac{\partial^2 U_{c0}}{\partial \sigma_i \partial \sigma_j} = b = -\frac{\nu}{E} \tag{3}$$

Thus, using the more common elastic parameters of Young's modulus ($E$) and Poisson's ratio ($\nu$), the elastic potential for linear elasticity is:

$$U_{c0} = U_0 = \frac{1}{2E}\left(\sigma_i^2 - 2\nu\sigma_i\sigma_j\right) \tag{4}$$

The elastic potential may also be formulated in terms of stress invariants or, a more general formulation, in terms of the 6 components of the stress tensor.

$$U_{c0} = U_0 = \frac{1+\nu}{2E}\left(\sigma_x^2 + \sigma_y^2 + \sigma_z^2 + 2\tau_{xy}^2 + 2\tau_{yz}^2 + 2\tau_{zx}^2\right) - \frac{\nu}{2E}\left(\sigma_x + \sigma_y + \sigma_z\right)^2 \tag{5}$$

It may be useful to decompose it in terms of spherical (volumetric) and deviatoric components:

$$U_{c0} = U_0 = U_v + U_d \tag{6}$$

where

$$U_v = \frac{1-2\nu}{6E}(\sigma_x + \sigma_y + \sigma_z)^2 \qquad (7)$$

and

$$U_d = \frac{1+\nu}{6E}\left[(\sigma_x - \sigma_y)^2 + (\sigma_y - \sigma_z)^2 + (\sigma_z - \sigma_x)^2 + 6(\tau_{xy}^2 + \tau_{yz}^2 + \tau_{zx}^2)\right] \qquad (8)$$

Using bulk and shear moduli ($K$ and $G$) and octahedral stresses, it may be expressed as

$$U_{c0} = U_0 = U_v + U_d = \frac{\sigma_{oct}^2}{2K} + \frac{3\tau_{oct}^2}{4G} \qquad (9)$$

Here, the formulation using principal stresses (Eq. 4) will mainly be used for the sake of simplicity.

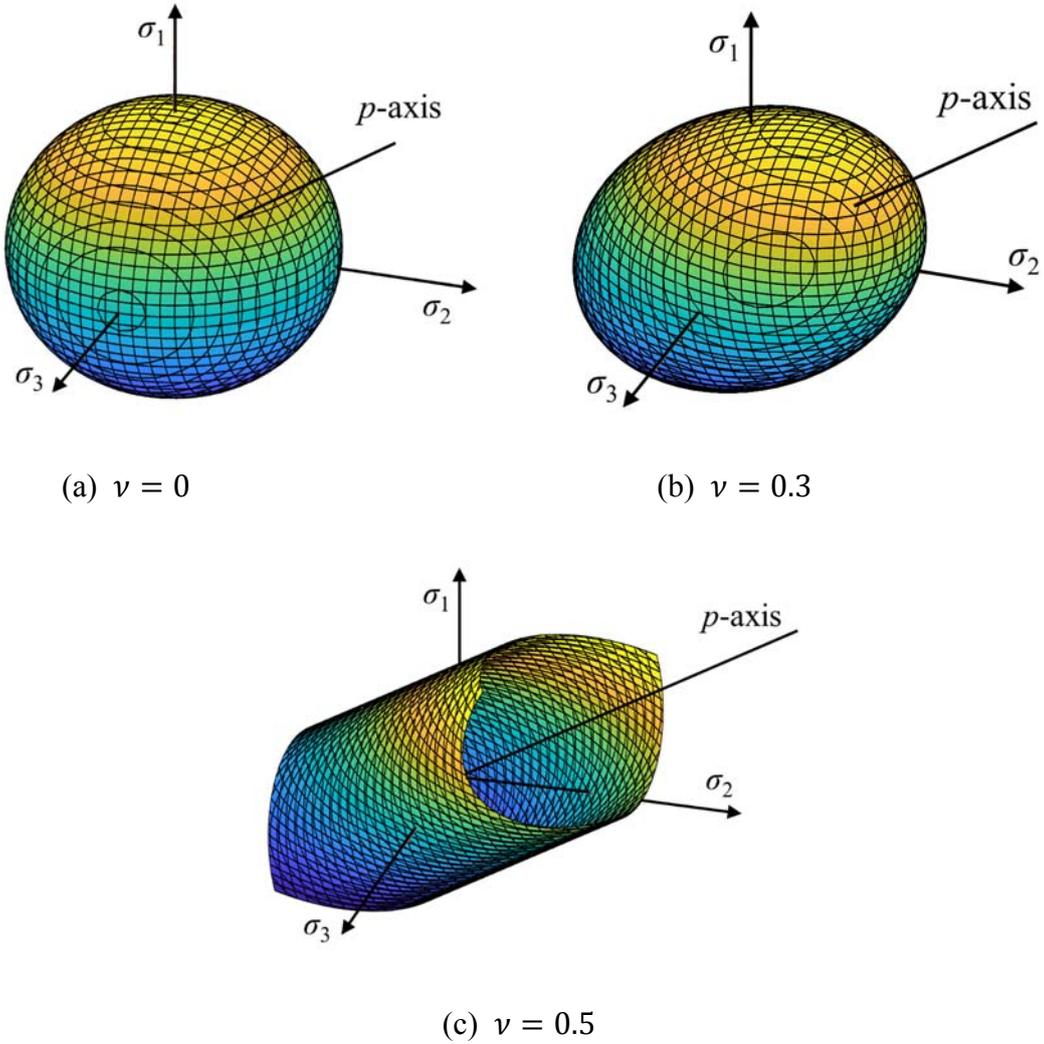

(a) $\nu = 0$

(b) $\nu = 0.3$

(c) $\nu = 0.5$

Figure 2. Linear elastic potentials in three-dimensional (3D) principal stress space: (a) $\nu = 0$ ; (b) $\nu = 0.3$ ; (c) $\nu = 0.5$ (von Mises).

The shape of the elastic potential in the principal stress space is an ellipsoid (Figure 2b). For the particular case of an incompressible material ($\nu = 0.5$), the elastic potential degenerates into a cylinder (Figure 2c); for the case of $\nu = 0$, it is a sphere (Figure 2a), and for the strange case of $\nu = -1$, it degenerates into two planar surfaces of maximum mean stress. The positive definite property of the strain energy ($U_{c0} \geq 0$) gives the limit values of the Poisson's ratio ($-1 \leq \nu \leq 1/2$). From a geometrical point of view, this means that the elastic potentials should be convex surfaces (Figure 2) (e.g. Callen 1985). The distance between the elastic potential and the origin gives the stiffness of the material in that direction. Thus, for example, $\nu = 0.5$ means that the stiffness in the mean stress direction is infinite, i.e. the bulk modulus is infinite ($K \to \infty$). For visualization purposes, the elastic potentials are also plotted in two dimensions in Figure 3, using the octahedral normal and shear stresses (Eq. 9).

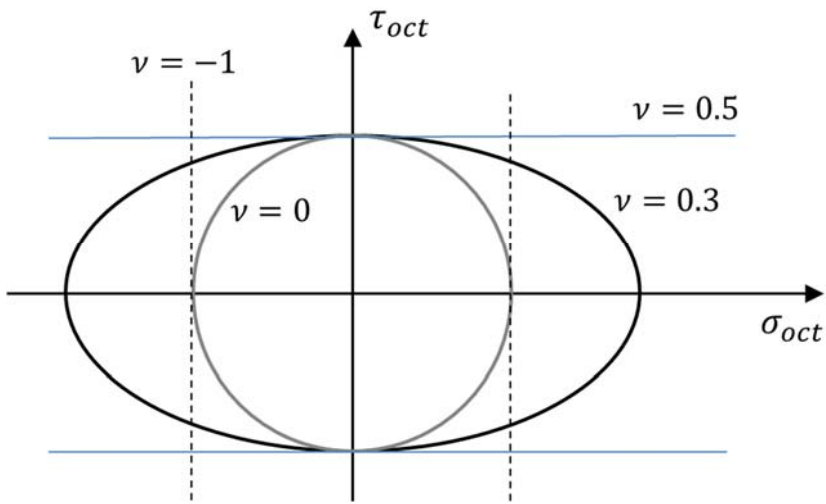

Figure 3. Linear elastic potentials for different Poisson's ratios in octahedral stress plot.

Using Eqs. (1) and (4), it may be shown that $\nu$ controls the strain path (shape of the elastic potential), while $1/2E$ acts like a kind of elastic multiplier:

$$\varepsilon_{ij}^e = \frac{\partial U_{c0}}{\partial \sigma_{ij}} = \frac{1}{2E}\frac{\partial (2EU_{c0})}{\partial \sigma_{ij}} \tag{10}$$

The elastic potential (Eq. 10) is somehow analogous to the plastic potential:

$$\dot{\varepsilon}_{ij}^p = \lambda \frac{\partial g}{\partial \sigma_{ij}} \tag{11}$$

where $\dot{\varepsilon}_{ij}^p$ is the plastic strain increment, $g$ is the plastic potential and $\lambda$ is the plastic multiplier. The only difference between Eqs. (10) and (11) in the mathematical formalism is that, in the plastic part (Eq. 11), the strains (and the plastic multiplier) are incrementally defined, while in the elastic part, they are absolute values.

## 2.2 Yield surface

The proposed approach assumes that the yield surface should correspond to an elastic potential surface for a critical value of the complementary strain energy density. In this way, an energetic criterion is used for the yield surface. For linear elasticity, $U_0$ and $U_{c0}$ are the same and it is not necessary to distinguish. However, the critical value should be defined in terms of $U_{c0}$ and not $U_0$ because of the principle of minimum complementary energy, which states that of all the elastic stress states satisfying the given boundary conditions, those that satisfy the equilibrium equations make the complementary energy a local minimum (e.g., Sadd 2014). Hence, it seems logical to impose the critical value to the property that constitutes a local minimum:

$$U_{c0,y} = U_{0,y} = \frac{1}{2E}\left(\sigma_i^2 - 2\nu \sigma_i \sigma_j\right) \tag{12}$$

This implies a yield surface that is an ellipsoid in the principal stress space (Figure 2). For the particular case of an incompressible material ($\nu = 0.5$), the yield surface is the von Mises (1913) cylinder (Figure 2c):

$$U_{c0,y} = U_{0,y} = \frac{\tau_y^2}{2G} \tag{13}$$

where $\tau_y$ is the yield stress of the material in pure shear and $G$ is the shear modulus.

## 2.3 Tension-compression yield asymmetry

Most materials show a tension-compression yield (and strength) asymmetry, i.e. the yield stress at compression is usually higher than that at tension. In Eq. (12), no distinction between compressive and tensile stresses was made because the predicted behaviour is symmetric at tension and compression. Besides, the quadratic form proposed in Eq. (2) for linear elasticity is not fully general because the linear term in stresses and the constant term, which are irrelevant for the linear elastic stress-strain behaviour, are missing.

$$U_{c0} = U_0 = a\sigma_i^2 + b\sigma_i\sigma_j + c\sigma_i + d \tag{14}$$

The strain energy density has units of energy per volume, i.e. pressure. Thus, coefficients $a$ and $b$ have units of the inverse of pressure (Eq. 3), $c$ is dimensionless and $d$ has units of pressure and it may be combined with the strain energy density for the sake of simplicity. Hence, Eq. (14) may be simplified by reorganizing the parameters as follows:

$$\sigma_i^2 + \frac{b}{a}\sigma_i\sigma_j + \frac{c}{a}\sigma_i + \frac{d-U_{c0}}{a} = 0 \tag{15}$$

where $b/a = -2\nu$ (Eq. 3). As for the Poisson's ratio, the values of the other parameters have certain limits to ensure that $U_{c0} \geq 0$.

The yield surface may be obtained imposing a limit value of the complementary strain energy density, $U_{c0,y}$ in Eq. (15). Once the Poisson's ratio is determined, the two remaining parameters may be obtained from the yield stresses for two different stress paths. For example, using the uniaxial tensile and compressive yield stresses ($-\sigma_t$ and $\sigma_c$), the yield surface is

$$\sigma_i^2 - 2\nu\sigma_i\sigma_j - (\sigma_c - \sigma_t)\sigma_i - \sigma_c\sigma_t = 0 \tag{16}$$

or using the hydrostatic tensile and compressive yield stresses ($-p_t$ and $p_c$), the yield surface is

$$\sigma_i^2 - 2\nu\sigma_i\sigma_j - (1-2\nu)(p_c - p_t)\sigma_i - 3(1-2\nu)p_c p_t = 0 \tag{17}$$

Thus, the yield surface has 3 parameters (e.g., $\nu$, $\sigma_c$ and $\sigma_t$, Eq. 16). Please, note that compressive stresses are assumed as positive and $\sigma_t$ and $p_t$ are positive (absolute) values.

The third term in Eqs. (14-17) causes a translation of the elastic potentials and the yield surface, which may be interpreted as a shifted origin or an initial hydrostatic stress state, $\sigma_0$, so that the elastic potential ellipsoid (Figure 2b) is shifted and its origin is at $\sigma_0$. Please, note that $\sigma_0$ is not an "apparent" or measurable initial stress and may be considered simply as a broad idealization of internal forces, stress history, atmospheric pressure.... Using $\sigma_0$, the elastic potential (Eq. 14) or the yield surface ($U_{c0,y}$) may be alternatively expressed as:

$$U_{c0} = U_0 = \frac{1}{2E}\left((\sigma_i - \sigma_0)^2 - 2\nu(\sigma_i - \sigma_0)(\sigma_j - \sigma_0)\right) \tag{18}$$

The relationship between $\sigma_0$ and coefficients $c$ and $d$ is given by Eqs. (14) and (18).

$$c = -\frac{1-2\nu}{E}\sigma_0 \text{ and } d = \frac{3(1-2\nu)}{2E}\sigma_0^2 \tag{19}$$

Using Eqs. (16-18), the initial or shifting stress ($\sigma_0$) may be expressed as a function of the yield stresses

$$\sigma_0 = \frac{\sigma_c - \sigma_t}{2(1-2\nu)} \tag{20}$$

$$\sigma_0 = \frac{p_c - p_t}{2} \tag{21}$$

For the particular case of $\nu=0.5$, the ellipsoid (Figure 2b) degenerates into the von Mises cylinder (Figure 2a), if tensile and compressive yield stresses are symmetric (e.g. $\sigma_c = \sigma_t$), or into an elliptic paraboloid, if they are not (e.g. $\sigma_c \neq \sigma_t$). For the latter case (paraboloid), the shifting stress $\sigma_0$ (Eq. 18) and $p_c$ (Eq. 17) are not applicable.

**2.4 Elliptical yield surfaces**

For compressible, linear materials, the elastic potential is an ellipsoid (Eq. 15) (Figure 2b) in a three-dimensional stress space and an ellipse in a two-dimensional stress plot (Figure 3). Without aiming to be comprehensive, there are some examples in the literature of elliptical yield surfaces that could be improved and better explained using the proposed approach:

- The yield surface of the Modified Cam Clay (MCC) model (Roscoe and Burland 1968) for soils. It does not consider the link between the shape of the ellipse and the Poisson's ratio. Appendix I shows the connection between the shape of the ellipse and the Poisson's ratio within the proposed framework for soils.

- The Ellipse failure criterion for metallic glasses (e.g., Liu et al. 2015). Compared to the proposed model, it does not consider the full 3D stress space and the shape at compression is changed "ad hoc" and is not convex. The application of the proposed linear model to metallic glasses is illustrated in Appendix II.

- The yield criterion for transversely isotropic solid foams by Ayyagari and Mural (2015). The formulation of its isotropic version coincides with the proposed approach, but the shifting stress ($\sigma_0$) is introduced in a "semi-phenomenological fashion".

## 2.5 Uniqueness of yield energy

The approach presented in this paper assumes that there is a unique complementary strain energy density for which the material yields ($U_{c0,y}$) (Eq. 12). While this may hold for some materials, e.g. materials with an amorphous or disordered structure such as soils and metallic glasses, it may not be valid for other materials. Consequently, the proposed approach may be considered just as a classifying criterion. For isotropic linear elastic materials (Eq. 12), it may be demonstrated that only two types of yield complementary strain energy at the most are possible. For practical purposes, this may be interpreted using a free parameter ($\nu_y$) instead of the Poisson's ratio, e.g. in Eq. 16

$$\sigma_i^2 - 2\nu_y \sigma_i \sigma_j - (\sigma_c - \sigma_t)\sigma_i - \sigma_c \sigma_t = 0 \tag{22}$$

The limits of $\nu_y$ are the same as those of the Poisson's ratio ($-1 \leq \nu_y \leq 0.5$), because the yield energy has to be positive. The value of $\nu_y$ determines the type of critical energy, e.g. $\nu_y = 0.5$ implies that the critical energy is just the distortional part (e.g. Hencky 1924). Interestingly, Eq. (22) with $\nu_y = 0.5$ is the elliptic paraboloid yield criterion proposed by Raghava et al. (1973) for polymers and used by Christensen (2013). Alternatively, the two types of yield energy may be treated independently and the yield criterion could be twofold (e.g. Christensen 2013).

## 3. Non-linear isotropic elasticity

### 3.1 A general example

Some materials, such as granular materials, show a non-linear response (Figure 1), even for the elastic range. In these stress-dependent materials, the stiffness is assumed to vary with the stress state.

For isotropic materials, the complementary strain energy density may be expressed just as a function of stress invariants. As for the linear case, the principal stresses will be here used for visualization of the elastic potential. Many different types of non-linear elasticity may be formulated within the hyperelastic framework (e.g., Humrickhouse et al. 2010; Houlsby and Puzrin 2006); here, for demonstration, the following complementary strain energy density function is assumed as an example:

$$U_{c0} = a\sigma_i^{2n} + b\sigma_i^n \sigma_j^n + c\sigma_i + d \qquad i,j,k = 1,2,3 \tag{23}$$

where $n$ is a material parameter that controls the material non-linearity. This formulation has the advantage that the stiffness is stress-dependent, not just mean pressure-dependent, and it may be reduce to the linear case (Eq. 14) by assuming $n=1$. Besides, the stiffness roughly follows a power law (approximately $E_i \propto \sigma_i^{2(1-n)}$). Thus, the common range is between $n=1$ (constant modulus) and $n=0.5$ (roughly linear stress-dependency of the stiffness). It is worth noting that non-linear hyperelastic models always introduce a "stress-induced" anisotropy (e.g., Puzrin, 2012). The analysis of the non-linear elastic behaviour of this hyperelastic model is detailed in Appendix III.

It is convenient to introduce two mathematical tweaks in Eq. (23). First, negative values of the stress are not possible in Eq. (23) when $n \neq 1$. Introducing a "back" stress, $\sigma_b$, (pressure) is useful to avoid negative values. Hence, positive "model" stress values, $\sigma^*$, are:

$$\sigma^* = \sigma + \sigma_b \tag{24}$$

This type of translation of the stress axes is quite common, for example, with the atmospheric pressure.

Secondly, it is useful to introduce a reference stress (pressure), $\sigma^*_{ref}$, so that the dimensions of constants $a$ and $b$ do not depend on $n$ and may be expressed as a function of a reference Young's modulus and a reference Poisson's ratio, $E_{ref}$ and $v_{ref}$, for that reference stress.

$\sigma_{ref}^*$ may be arbitrarily chosen, but $\sigma_b$ is a fitting parameter that determines the stress for which the stiffness is null. Thus, Eq. (23) may be expressed as:

$$U_{c0} = \frac{\sigma_{ref}^{*2}}{E_{ref}} \frac{1}{2n(2n-1)} \left(\frac{\sigma_i^*}{\sigma_{ref}^*}\right)^{2n} - v_{ref} \frac{\sigma_{ref}^{*2}}{E_{ref}} \frac{1}{n^2} \left(\frac{\sigma_i^*}{\sigma_{ref}^*}\right)^n \left(\frac{\sigma_j^*}{\sigma_{ref}^*}\right)^n + c \frac{\sigma_i^*}{\sigma_{ref}^*} + d \qquad (25)$$

Similarly to the linear elastic case, the yield surface may be defined as the elastic potential (Eq. 25) for a limit value of the complementary strain energy density ($U_{c0,y}$). Consequently, once the non-linear elastic constants ($E_{ref}, v_{ref}, n, \sigma_b$) have been determined, the additional two constants of the yield surface ($c, U_{c0,y} - d$) may be obtained from the yield stresses for two different stress paths, for example, the uniaxial or hydrostatic yield stresses at tension and compression.

For the non-linear case, the shape of the elastic potentials and the yield surface in the principal stress space are distorted ellipsoids (Figures 4-7). In Figures 4-7, simple values have been chosen for the constants, namely $E_{ref}=\sigma_{ref}^*=1$ (arbitrary units). The non-linear elastic potentials reflect the asymmetries caused by the non-linear elastic behaviour, such as larger elastic regions for higher compressive stresses, both in the triaxial and deviatoric planes (Figures 6 and 7, respectively). Hence, the yield surface is larger in those directions where the material stiffness is larger. In the deviatoric plane, the stress-dependency (non-linearity) distorts the circular section for linear elastic materials ($n=1$) towards rounded triangles, similar to Lade-Duncan (1974) or Matsuoka-Nakai (1974) surfaces, which in turn may be viewed as a kind of rounded Mohr-Coulomb.

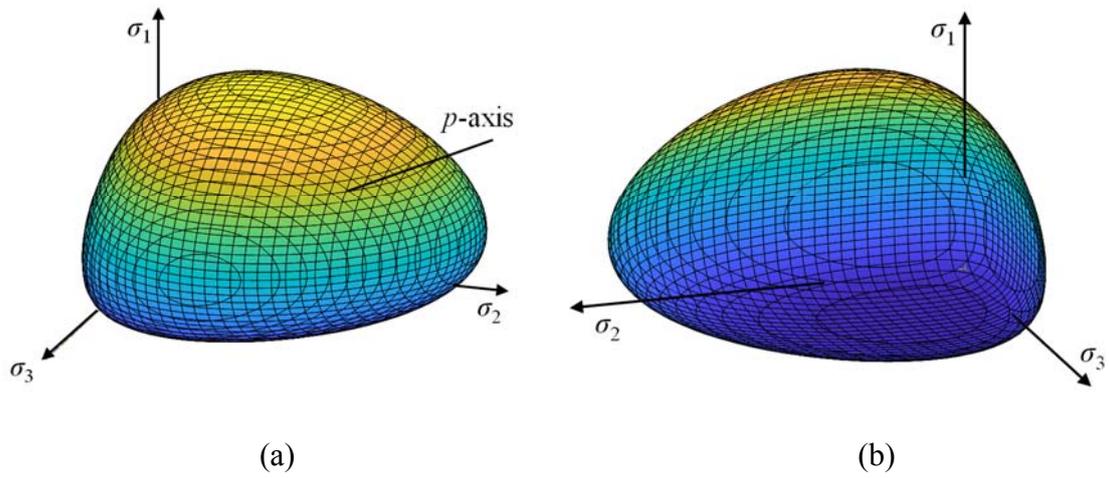

(a) (b)

Figure 4. Non-linear elastic potential in 3D principal stress space ($n \approx 0.5$, $v_{ref}=0.3$, $p_t=0$, $\sigma_b=0.1p_c$): (a) compressive side view; (b) tensile side view.

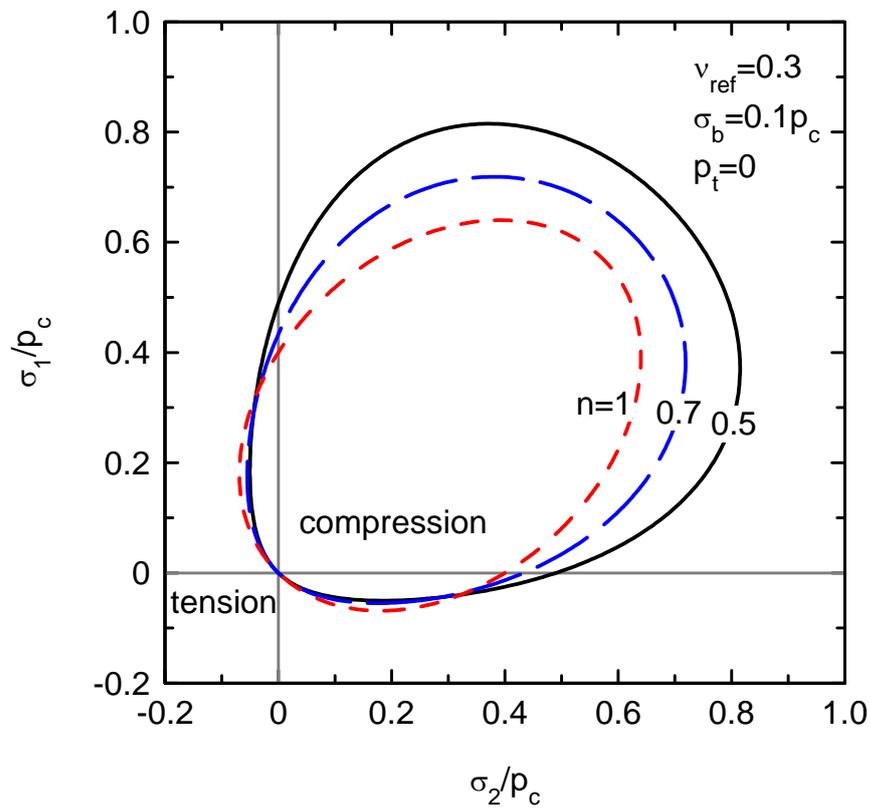

Figure 5. Biaxial non-linear elastic potentials ($\sigma_3 = 0$).

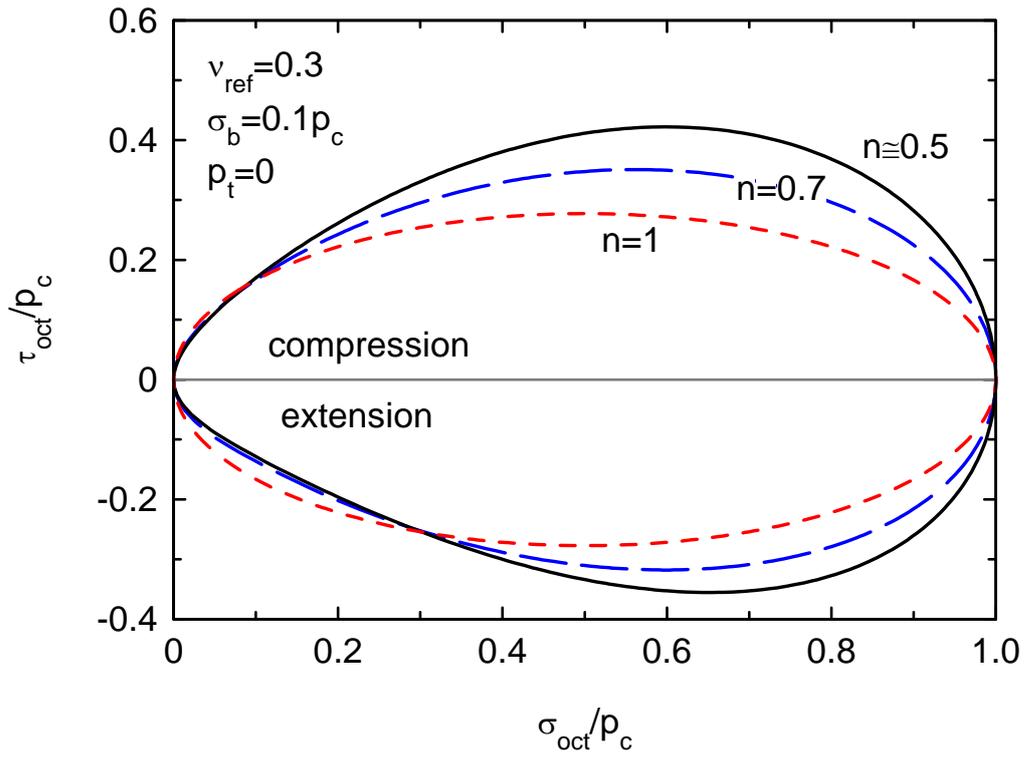

Figure 6. Non-linear elastic potentials in octahedral stress plot for the triaxial plane ($\sigma_2 = \sigma_3$).

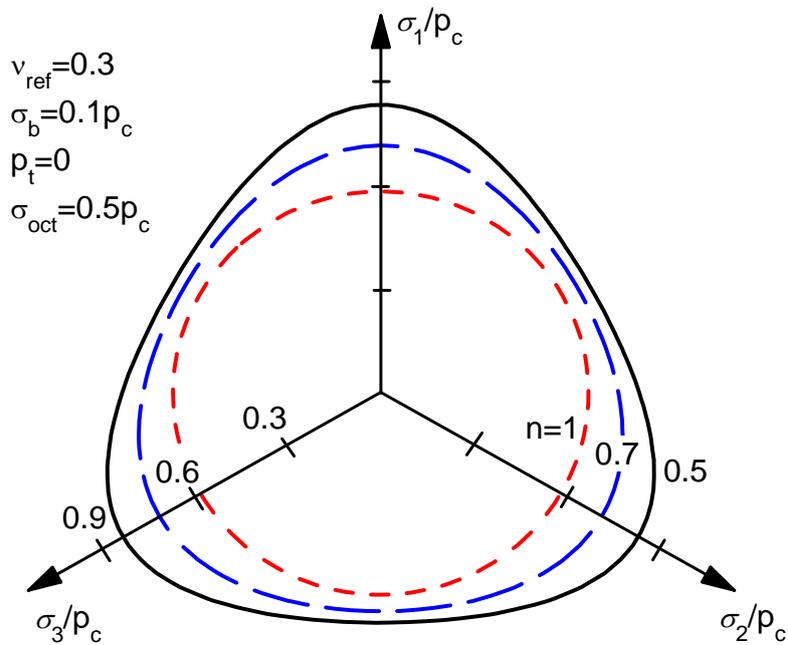

Figure 7. Non-linear elastic potentials in deviatoric plane.

The selection of appropriate hyperelastic models and fitting of yielding data for specific materials is far beyond the scope of this paper, but distorted ellipses as those in Figure 6 are used in the literature as yield surfaces (e.g., Bigoni and Piccolroaz 2004).

**3.2 Incompressible non-linear isotropic materials**

A simple way of formulating a non-linear incompressible hyperelastic model is by using just the deviatoric component of the complementary strain energy density (Eq. 8):

$$U_{c0} = U_d = \frac{1}{12G_{ref}}\left[\left(\frac{\sigma_1-\sigma_2}{\sigma_{ref}}\right)^{2n} + \left(\frac{\sigma_2-\sigma_3}{\sigma_{ref}}\right)^{2n} + \left(\frac{\sigma_1-\sigma_3}{\sigma_{ref}}\right)^{2n}\right] \qquad (26)$$

Similarly to Eq. (22), the quadratic power of 2 is replaced by $2n$ in Eq. (26). Besides, ordered principal stresses are considered in Eq. (26) to avoid negative values in the base of the $2n$ exponent. For full visualization in the principal stress space, they may be alternated. Assuming $\sigma_{ref}=1$ (arbitrary units) for the sake of simplicity, the yield surface is:

$$12G_{ref}U_{c0,y} = (\sigma_1 - \sigma_2)^{2n} + (\sigma_2 - \sigma_3)^{2n} + (\sigma_1 - \sigma_3)^{2n} \qquad (27)$$

Eq. (27) is equivalent to the Hosford (1972) yield criterion. The results for several $n$ values are plotted in the deviatoric or $\pi$–plane (Figure 8) and they do not vary with the mean pressure. Interestingly, the yield surface reduces to the von Mises criterion for the linear case ($n=1$) (as already mentioned) and to the Tresca criterion for a linear stress-dependency ($n=0.5$). The information available in the literature (e.g., Lade 1988) confirms that the yielding of non-linear incompressible materials, such as soft soils under undrained conditions, is better captured by Tresca than by von Mises criterion.

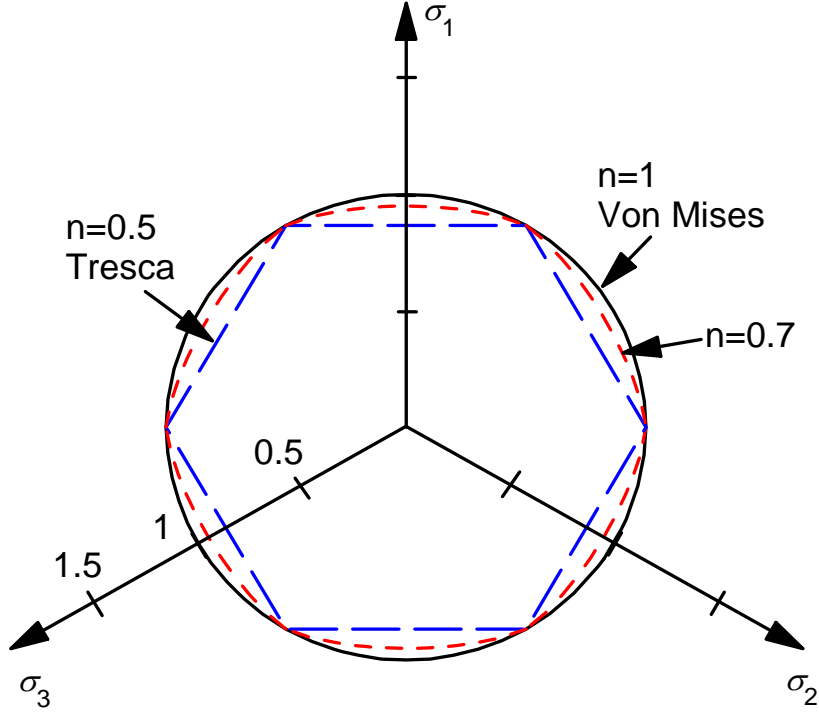

Figure 8. Influence of non-linearity for incompressible materials: deviatoric plane section of the yield surface.

## 4. Linear anisotropic elasticity

### 4.1 Transverse isotropic materials

As for the previous examples, the application of the proposed approach to anisotropic elasticity depends on the selected hyperelastic model. For the sake of simplicity, the case of transversely isotropic materials is considered, i.e. an axis of symmetry and only 5 independent elastic constants. Besides, the symmetry or longitudinal axis is assumed to be the vertical axis (equal to Axis 1), while the transverse axes are assumed as the horizontal ones (Axes 2 and 3). Thus, the (complementary) strain energy density is

$$U_{c0} = U_0 = \frac{1}{2E_V}\sigma_1^2 + \frac{1}{2E_H}(\sigma_2^2 + \sigma_3^2) - \frac{\nu_{HV}}{E_V}\sigma_1(\sigma_2 + \sigma_3) - \frac{\nu_{HH}}{E_H}\sigma_2\sigma_3 + c_V\sigma_1 +$$

$$c_H(\sigma_2 + \sigma_3) + d \qquad (28)$$

where $c_V$ and $c_H$ are the vertical and horizontal parameters that do not affect the stiffness matrix but cause a translation of the elastic potential (similarly to $c$ in Eq. 14). Using octahedral stresses as in Eq. 9 (Figure 3), an additional term that causes a rotation of the ellipse appears

$$U_{c0} = U_0 = U_v + U_d = \frac{(\sigma_{oct}-\sigma_0)^2}{2K} + \frac{3(\tau_{oct}-\tau_0)^2}{4G} + \frac{(\sigma_{oct}-\sigma_0)(\tau_{oct}-\tau_0)}{J} \qquad (29)$$

where $K$, $G$ and $J$ may be related to the 5 elastic parameters in Eq. (28) and $\sigma_0$ and $\tau_0$ give the translated origin and may be related to $c_V$ and $c_H$.

Following the approach used above, the yield surface can be found by imposing a limit value of the complementary strain energy density ($U_{c0,y}$). However, if the material is, for example, fibre reinforced, it is clear that the yield energy in the longitudinal (vertical) direction is different from that in the transverse direction and the yield parameters (e.g. $J$ in the corresponding yield surface from Eq. 29) may not be directly linked to the elastic parameters.

**4.2 Incompressible linear anisotropic materials**

As in Section 3.2, the complementary strain energy for incompressible materials may be formulated using just the distortional part. For anisotropic materials, the shear moduli in the different directions have to be considered. As in the previous Section 4.1, the axes of anisotropy are assumed to coincide with the principal stresses. Thus, the (complementary) strain energy is

$$U_{c0} = U_d = \frac{(\sigma_1-\sigma_2)^2}{12G_{12}} + \frac{(\sigma_2-\sigma_3)^2}{12G_{23}} + \frac{(\sigma_3-\sigma_1)^2}{12G_{31}} \qquad (30)$$

If different yield energies are assumed for the three planes of anisotropy ($U_{c0,yij}$), the classical Hill (1948) yield criterion is obtained.

$$1 = \frac{1}{U_{c0,y12}} \frac{(\sigma_1-\sigma_2)^2}{12G_{12}} + \frac{1}{U_{c0,y23}} \frac{(\sigma_2-\sigma_3)^2}{12G_{23}} + \frac{1}{U_{c0,y31}} \frac{(\sigma_3-\sigma_1)^2}{12G_{31}} \qquad (31)$$

**Conclusions**

This paper shows that yield surfaces may be assumed to be elastic potential surfaces for specific levels of critical complementary strain energy density. Traditional approaches, such as the total strain energy criterion, only consider second order terms, i.e. the initial strain energy is null and the elastic potential is centred at the current stress state. Here, first order terms are considered, and consequently, the elastic potential may be translated. The proposed approach shows a correlation between the shape of the yield surface and the Poisson's ratio, which control the shape of the elastic potential. This correlation agrees well with published values in the literature for soils and metallic glasses.

Introducing non-linear elasticity gives a wide range of elastic potentials, such as distorted ellipsoids or cylinders, with similar or identical shapes as previously published yield surfaces.

As hyperelasticity or associated plasticity, the proposed framework to derive associated or hyper yield surfaces using elastic potentials may be considered just as a classifying criterion and a possible approach to formulate yield surfaces.

**Acknowledgements**

The author would like to express his gratitude to the Spanish Ministry of Economy and Competitiveness and to the European Regional Development Fund (ERDF) for financing the National Plan Project (Ref.: BIA2015-67479-R) under the name of 'The Critical Distance in Rock Fracture'.

# Appendix I. Application of the linear isotropic case to soils

Soil response is clearly non-linear, but the application of the linear formulation (Section 2) gives a first approximation, as will be shown. As commonly done for soils, compressive stresses are assumed to be positive and the stress invariants $p$ and $q$ are used:

$$q = \sqrt{\tfrac{1}{2}[(\sigma_1-\sigma_2)^2 + (\sigma_2-\sigma_3)^2 + (\sigma_3-\sigma_1)^2]} \ ; \ p = \tfrac{\sigma_1+\sigma_2+\sigma_3}{3} \tag{AI.1}$$

Thus, using those stress invariants ($p$ and $q$) and $K$ and $G$, the yield surface based on the elastic potential (Eqs. 4, 9 or 12) may be expressed as

$$U_{c0,y} = \frac{q^2}{6G} + \frac{p^2}{2K} \tag{AI.2}$$

It is necessary to apply the translation to account for the tension-compression asymmetry (Eq. 18). Besides, in the case of soils, tensile stresses are usually null ($p_t \approx 0$) and the hydrostatic compressive yielding stress ($p_c$) is usually called the mean preconsolidation pressure. Hence, the translation given by Eq. (21) is $p_c/2$ and Eq. (AI.2) becomes

$$U_{c0,y} = \frac{q^2}{6G} + \frac{(p-p_c/2)^2}{2K} \tag{AI.3}$$

The shape of the elastic potential and yield surface given by Eq. (AI.3) is completely analogous to the yield surface and plastic potential of the MCC model (Roscoe and Burland, 1968)

$$f = g = q^2 + M^2 p(p - p_c) = 0 \tag{AI.4}$$

where $M$ is the stress ratio at critical state and may be correlated with the critical state friction angle for triaxial compression ($\sigma_2 = \sigma_3$, ordered principal stresses) as follows

$$sin\phi_{cr} = \frac{3M}{6+M} \tag{AI.5}$$

Thus, the analogy between the proposed yield surface (Eq. AI.3) and that of the MCC model (AI.4) gives the following equivalences:

$$M^2 = \frac{3G}{K} = \frac{9(1-2\nu)}{2(1+\nu)} \tag{AI.6}$$

$$p_c^2 = 8KU_{c0,y} \tag{AI.7}$$

Eq. (AI.6) implies a direct relationship between the Poisson's ratio (an elastic parameter of the soil) and the stress ratio at critical state (a plastic and failure parameter of the soil). The relationship is plotted in Figure AI.1 and generally agrees with the published values (Table AI.1). The published values correspond to parameters for specific constitutive models calibrated from laboratory experiments, mainly drained triaxial compression tests. Some scatter in the data may arise from soil anisotropy, soil nonlinearity and Poisson's ratio determination, calibration or specific meaning within the used constitutive model.

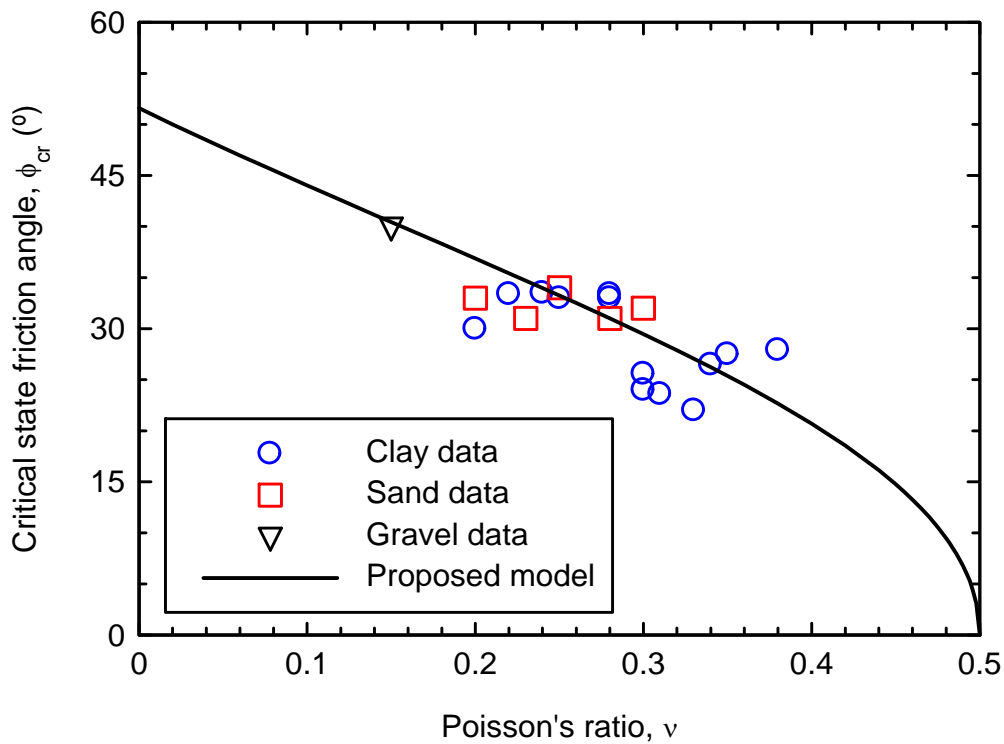

Figure AI.1. Relationship between Poisson's ratio and critical friction angle for soils.

Table AI.1. Published values of Poisson's ratio and critical friction angle.

| Soil | $\nu$ | $\phi_{cr}(°)$ | Reference |
|---|---|---|---|
| Beaucaire Marl (reconstituted silty clay) | 0.25 | 33.0 | Masin et al. (2006) |
| Boston blue clay | 0.24 | 33.5 | Papadimitriou et al. (2005) |
| Boston blue clay | 0.28 | 33.4 | Whittle and Satabutr (2005) |
| Brno clay | 0.33 | 22 | Masin (2013) |
| Dortmund clay | 0.38 | 27.9 | Masin (2013) |
| Empire clay | 0.31 | 23.6 | Whittle and Satabutr (2005) |
| Kaolin clay | 0.34 | 26.5 | Castro et al. (2013) |
| Kaolin clay | 0.35 | 27.5 | Masin (2013) |
| Koper clay | 0.28 | 33 | Masin (2013) |
| Lower Cromer Till | 0.20 | 30 | Papadimitriou et al. (2005) |
| Mexico Gulf clay | 0.30 | 25.6 | Whittle and Satabutr (2005) |
| Norrköping clay | 0.22 | 33.4 | Rouainia and Wood (2000) |
| Weald clay | 0.30 | 24 | Masin (2013) |
| Berlin sand | 0.28 | 31 | Pestana et al. (2005) |
| Hostun sand | 0.20 | 33 | Chang and Hicher (2005) |
| Sacramento river sand | 0.25 | 34 | Wan and Guo (1998) |
| Toyura sand | 0.23 | 31 | Pestana et al. (2005) |
| Toyura sand | 0.30 | 32 | Zhang et al. (2010) |
| Calcareous fine gravel | 0.15 | 40 | Castro et al. (2013) |

References in Table AI.1

## Appendix II. Application of the linear isotropic case to metallic glasses

Based on experiments in metallic glasses, Zhang and Eckert (2005) proposed an Ellipse failure criterion. Furthermore, Liu et al. (2015) found a very interesting correlation between the shape of the ellipse and the Poisson's ratio. The advantages of metallic glasses for this study are that they are macroscopically isotropic and homogeneous, exhibit nearly zero tensile ductility and very limited compressive plasticity, and cannot be work-hardened. Thus, the yield surface may be assumed to be the failure surface. Here, the data gathered by Liu et al. (2015) are reinterpreted within the proposed framework for linear elastic isotropic materials (Section 2).

For the sake of comparison with Liu et al. (2015), the yield surface (Eq. 18) is represented as a shifted ellipse in the Mohr's diagram (normal and shear stress on the failure plane $(\sigma, \tau)$):

$$\left(\frac{\tau}{\tau_y}\right)^2 + \left(\frac{\sigma - \sigma_0}{\sigma_y}\right)^2 = 1 \tag{AII.1}$$

where $\sigma_0$ is the initial (or shifting) stress and $\tau_y$ and $\sigma_y$ are the vertical and horizontal semi-axes of the ellipse, respectively. Their ratio is the parameter that controls the shape of the ellipse, $\alpha = \tau_y/\sigma_y$, and may be related to the Poisson's ratio, $\nu$. For the sake of consistency with other parts of this paper, compressive stresses are assumed to be positive. The main differences between the proposed approach for the linear case and the Ellipse criterion (e.g., Liu et al. 2015) are that the tension-compression asymmetry is introduced through the shifting stress $\sigma_0$ and that the proposed yield surface is convex and fully 3D. Chen et al. (2011) proposed a 2D eccentric (shifted) ellipse, whose derivation is different from the present approach, but it provides a failure envelope as that in Eq. AII.1.

From Eq. (18), using the Mohr's circle and assuming triaxial ($\sigma_2 = \sigma_3$) or plane strain ($\sigma_2 = \nu(\sigma_1 + \sigma_3)$) conditions, the following relationships between $\alpha$ and $\nu$ may be found:

$$\alpha^2 = \frac{1-2\nu}{2} \qquad \text{Triaxial} \tag{AII.2}$$

$$\alpha^2 = \frac{1-2\nu}{2(1-\nu)} \qquad \text{Plane strain} \tag{AII.3}$$

Figure AII.1 shows the relationship between $\alpha$ and $\nu$. The correlation is equivalent to that shown in Figure AI.1 for soils. Although some uncertainties arise in the comparison because the stress triaxiality of the data is not clear and experimental $\alpha$ values are influenced by their calculation process, the correlation between the shape of the yield surface and the Poisson's ratio is clear.

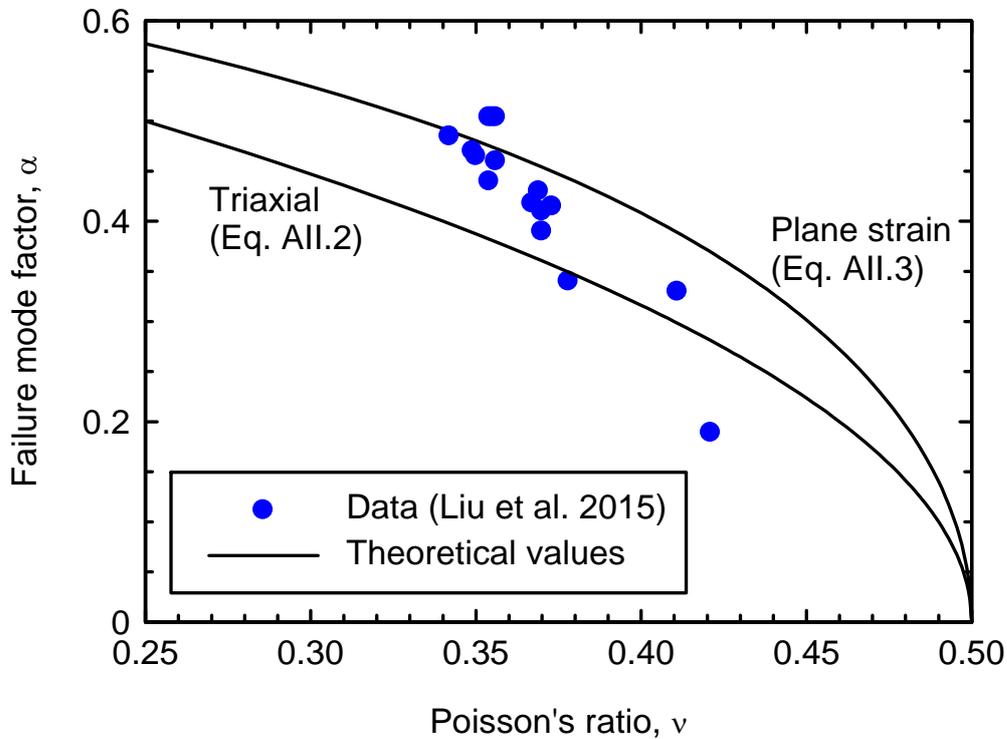

Figure AII.1. Relationship between $\alpha$ and $\nu$ for metallic glasses.

Similarly to Eq. (20), the shifting stress $\sigma_0$ may be determined based on the uniaxial compression and tensile strengths ($\sigma_c$ and $-\sigma_t$, respectively):

$$\sigma_0 = \frac{\sigma_c - \sigma_t}{4\alpha^2} \qquad \text{(AII.4)}$$

As an example of the matching properties of the proposed yield surface (Eq. AII.1), the data by Qu et al. (2011) are fitted in Figure AII.2. The fitting is based on the tensile and compressive strengths ($\sigma_c = 1.84$ GPa and $\sigma_t = 1.66$ GPa) and the $\alpha$ value given by Qu et al. (2011) ($\alpha=0.41$), which is in the range provided by triaxial (Eq. AII.2, $\alpha=0.36$) and plane strain (Eq. AII.3, $\alpha=0.45$) conditions using the reported value of the Poisson's ratio ($\nu = 0.37$) (Liu et al. 2015).

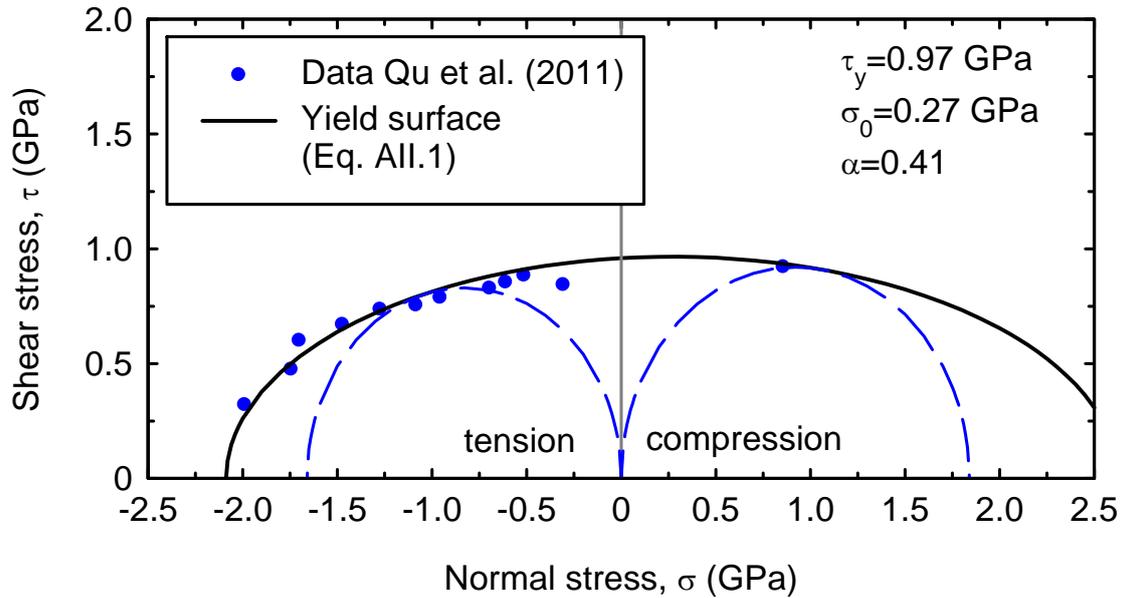

Figure AII.2. Fitted yield surface for data of a metallic glass tested by Qu et al. (2011).

## Appendix III. Example of non-linear hyperelastic model

This appendix presents the elastic behaviour, specifically the variation of the stiffness with the stress level, that corresponds to the complementary strain energy density function used as an example in Section 3 (Eq. 25). For simplicity, a null "back" stress is assumed ($\sigma_b = 0$), so, $\sigma^* = \sigma$. Otherwise, the corresponding isotropic stress translation must be applied.

Using equivalent moduli, the symmetric compliance matrix ($C_{ij}$) may be expressed as follows:

$$\begin{bmatrix} \varepsilon_1 \\ \varepsilon_2 \\ \varepsilon_3 \end{bmatrix} = \begin{bmatrix} \frac{1}{E_1} & -\frac{\nu_{12}}{E_{12}} & -\frac{\nu_{13}}{E_{13}} \\ -\frac{\nu_{12}}{E_{12}} & \frac{1}{E_2} & -\frac{\nu_{23}}{E_{23}} \\ -\frac{\nu_{13}}{E_{13}} & -\frac{\nu_{23}}{E_{23}} & \frac{1}{E_3} \end{bmatrix} \begin{bmatrix} \sigma_1 \\ \sigma_2 \\ \sigma_3 \end{bmatrix} \qquad (AIII.1)$$

where the equivalent moduli may be derived from the complementary strain energy density function.

$$C_{ij} = \frac{\partial^2 U_{c0}}{\partial \sigma_i \partial \sigma_j} \qquad (AIII.2)$$

Thus, their values are

$$C_{11} = \frac{1}{E_1} = \frac{1}{E_{ref}} \left( \frac{\sigma_1}{\sigma_{ref}} \right)^{2(n-1)} - \frac{\nu_{ref}}{E_{ref}} \frac{n-1}{n} \left( \frac{\sigma_1}{\sigma_{ref}} \right)^{n-2} \frac{\sigma_2^n + \sigma_3^n}{\sigma_{ref}^n} \qquad (AIII.3)$$

$$C_{12} = \frac{\nu_{12}}{E_{12}} = \frac{\nu_{ref}}{E_{ref}} \left( \frac{\sigma_1}{\sigma_{ref}} \right)^{n-1} \left( \frac{\sigma_2}{\sigma_{ref}} \right)^{n-1} \qquad (AIII.4)$$

Figure AIII.1 shows the variation of the uniaxial stiffness with the corresponding principal stress (Eq. AIII.3).

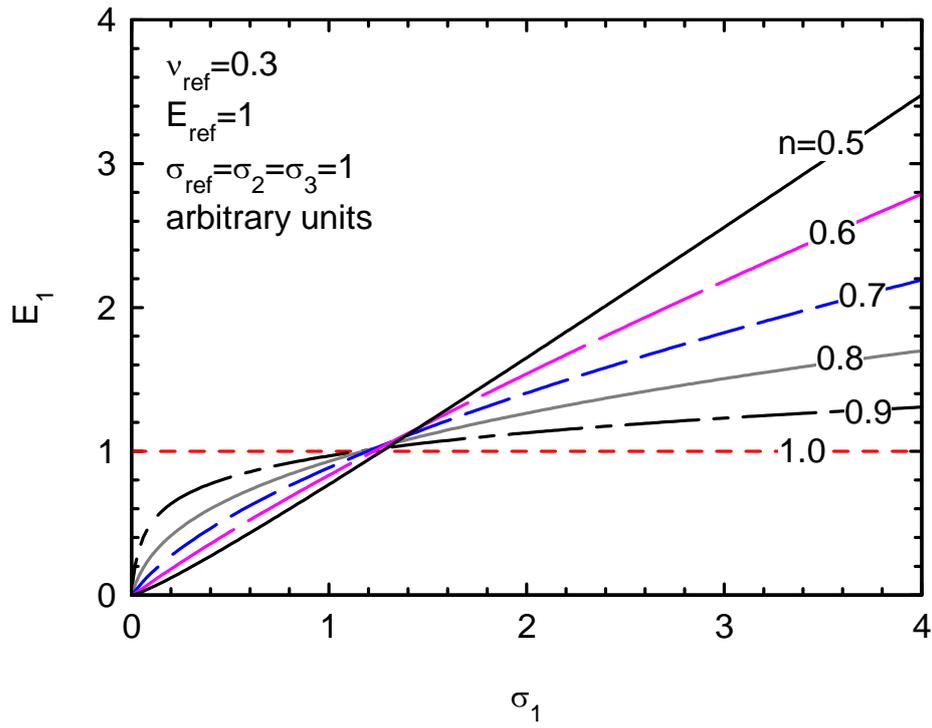

Figure AIII.1. Stiffness variation with the stress level.

# List of symbols

| | |
|---|---|
| *a,b,c,d* | Parameters of the complementary strain energy function |
| $C_{ij}$ | Compliance matrix |
| *E* | Young's modulus |
| *f* | Yield surface |
| *g* | Plastic potential |
| *G* | Shear modulus |
| *K* | Bulk modulus |
| *M* | Stress ratio at critical state |
| *n* | Non-linearity material parameter |
| *p* | Mean stress |
| $p_c$ | Hydrostatic compressive yield stress |
| $p_t$ | Hydrostatic tensile yield stress |
| *q* | Deviatoric stress |
| $U_0$ | Strain energy density |
| $U_{c0}$ | Complementary strain energy density |
| *α* | Shape factor in the Ellipse failure criterion |
| *ε* | Strain |
| $\varepsilon_{ij}$ | Strain tensor component |
| *λ* | Plastic multiplier |
| *ν* | Poisson's ratio |
| *σ* | Normal stress |
| $\sigma^*$ | Positive "model" stress |

$\sigma_b$     "Back" stress

$\sigma_c$     Uniaxial compressive yield stress

$\sigma_i$     Principal stress

$\sigma_t$     Uniaxial tensile yield stress

$\sigma_0$     Initial or shifting stress

$\sigma_{ij}$     Stress tensor component

$\tau$     Shear stress

$\phi_{cr}$     Critical friction angle

Subscripts/superscripts:

| | |
|---|---|
| 0 | Initial |
| 1,2,3 | unordered principal stress directions |
| *d,v* | deviatoric, volumetric |
| *e,p* | elastic, plastic |
| *i,j,k* | principal stress directions in contracted notation |
| *oct* | octahedral |
| *ref* | reference |
| *x,y,z* | coordinate axes |
| *y* | yield |

Compressive stresses and strains are assumed to be positive.